\documentstyle[12pt]{article}

         \def\la{\lambda}
         \def\be{\begin{equation}}
         \def\bea{\begin{eqnarray}}

         \def\ep{\epsilon}

         \def\ee{\end{equation}}
         \def\eea{\end{eqnarray}}
         \def\R{\rm {I\kern-.200em R}}
         \def\C{\rm {I\kern-.520em C}}
          \def\s{\sigma }
\begin{document}
\begin{titlepage}
\begin{center} {\Large \bf Exact solution of a one-parameter family of\\
asymmetric exclusion processes}\\
\centerline {\bf M. Alimohammadi$ ^{a,c}$ \footnote
{e-mail:alimohmd@netware2.ipm.ac.ir} ,
V. Karimipour$ ^{b,c}$ \footnote {e-mail:vahid@netware2.ipm.ac.ir},
M. Khorrami$ ^{b,c,d}$ \footnote {e-mail:mamwad@netware2.ipm.ac.ir}}
\vskip 1cm
{\it $^a$ Department of Physics, University of Teheran, North Karegar,
} \\
{\it Tehran, Iran }\\
{\it $^b$ Department of Physics, Sharif University of Technology, }\\
{\it P.O.Box 11365-9161, Tehran, Iran }\\
{\it $^c$ Institute for Studies in Theoretical Physics and Mathematics,
}\\
{\it P.O.Box 19395-5746, Tehran, Iran}\\
{\it $^d$ Institute for Advanced Studies in Basic Physics , P.O.Box 159
159, }\\
{\it  Gava Zang , Zanjan 45195 , Iran }\\
\end{center}

\begin{abstract}

We define a family of asymmetric processes for particles on a
one-dimensional
lattice, depending on a continuous parameter $\lambda \in [0,1] $,
interpolating between the completely asymmetric processes [1] (for
$\lambda =1 $ ) and the $n=1$ drop-push models [2] (for $ \lambda =0$ ).
For arbitrary $\la$, the model describes an exclusion process, in
which a particle pushes
its right neighbouring particles to the right, with rates depending
on the number of these particles. Using
the Bethe ansatz, we obtain the exact solution of the master equation .
\end{abstract}


\end{titlepage}
\newpage
\section{Introduction}
Various versions of one-dimensional asymmetric simple exclusion
processes (ASEP) have been shown to be of physical interest in a
variety of problems including the Kinetics of biopolymerization [3],
polymers in random media, dynamical models
of interface growth [4], and traffic models [5]. This model is also
related to the
noisy Burgers equation [6], and thus to the study of shocks [7,8].
Besides the equilibrium properties of this model, its dynamical
properties have also been studied in [8-10].

Recently the totally  ASEP model, with sequencial updating on an
infinite lattice, has been solved
 exactly by G. M. Sch\"utz [1] using the coordinate Bethe ansatz.  In
this model, each lattice site can be occupied by at most one particle
and a particle hops with rate one to its right
 neighbouring site if it is not already occupied; otherwise
the attempted move is rejected. In his work, instead of using
the quantum Hamiltonian formalism, which is suitable for studying
the dynamical
exponents and certain time-dependent correlation functions, Sch\"utz
adopted the coordinate representation for writing the master equation. By
solving the master equation exactly, he was able to obtain explicit
expressions for conditional probabilities
 $P(x_1,x_2,\cdots ,x_N;t|y_1,y_2,\cdots ,y_N;0)$ of finding
 $N$ particles on lattice sites $x_1,\cdots x_N$ at time $t$ with
initial occupation $y_1,\cdots ,y_N$ at time $t=0$.

The master equation for the probability of finding particle 1 on site
$ k_1$ , particle 2 on site
 $ k_2 , \cdots,$ and particle $N$ on site $ k_N $, with
$ k_N > k_{N-1} > ...k_2 > k_1 $, is written as
$$ {\partial\over \partial t}  P( k_1, k_2,\cdots ,k_N,t) = P( k_1-1 , k_2,
\cdots ,k_N,t ) + P( k_1, k_2-1,\cdots ,k_N,t ) +\cdots $$
\be +P( k_1, k_2,\cdots ,
k_N-1,t ) - N P( k_1, k_2,\cdots , k_N,t ), \ee
if $k_{i+1}-k_i>1$. This equation was then augmented by the following
boundary condition
\be P(k,k,t) = P(k,k+1,t), \hskip 1cm \forall k.  \ee
In writing (2), we have supressed for simplicity the position of all
the other particles,
bearing in mind that this condition should hold for every pair of
adjacent variables $k_i$ and $ k_{i+1} $. In the following we always
use this simplified notation. It was then shown
 that (1) and (2) give the correct master equation in the whole physical
region ( i.e. the region $ k_i < k_{i+1} $ ) for the probabilities.
In the rest of [1], the exact solution of the master equation (1)
(with boundary condition (2)) is constructed.

We now describe what we have done in the present paper. In section 2,
we substitute the boundary condition (2) by
\be P(k,k,t) = P(k-1,k,t), \hskip 1cm \forall k, \ee
and show that this boundary condition, together with (1), describes the
$n=1$ drop-push dynamics [2]. In this process, even if the right
neighbouring sites of a particle are occupied,
the particle hops with rate one to the next right site, pushing the right
 neighbouring particles to the next sites. This means that all
the following processes occur with equal rate one :
$$ A0 \rightarrow 0A, 
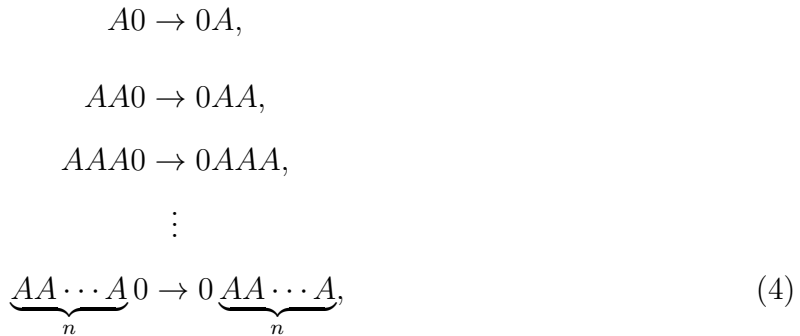
$$
$$ AA0 \rightarrow 0AA, $$
$$ AAA0 \rightarrow 0AAA, $$
$$ \vdots $$
\be \underbrace {AA\cdots A}_n0 \rightarrow 0
\underbrace {AA\cdots A}_n, \ee
where we have adopted the standard notation for representing a particle
by $A$ and a vacancy by $0$. We then obtain a closed form for the
conditional proabilities for this process.

This process, in which a particle pushes as many particles with rate
one, is the opposite extreme of what was solved by Sch\"utz, and
interestingly admits a closed form solution for the conditional
probabilities $P(x_1,x_2,\cdots ,x_N;t|y_1,y_2,\cdots ,y_N;0)$ in the
form of an $N \times N$ determinant.

In section 3, we combine the boundary condition (2) and (3) in the form
\be P(k,k,t) =\lambda P(k,k+1,t)+(1-\lambda) P(k-1,k,t), \hskip 1cm
\forall k,  \ee
and show that the resulting master equation ((1) and (5) ), describes
a processin which the
processes shown in (4) occur with unequal rates: namely the process
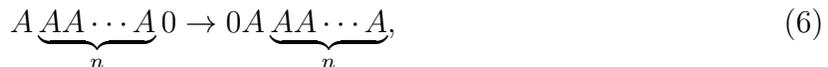
\be A \underbrace {AA\cdots A}_n0 \rightarrow 0 A\underbrace
{AA\cdots A}_n, \ee
occurs with rate
\be r_n={1 \over {1+{\lambda \over \mu } +({\lambda \over \mu } )^2
+\cdots+({\lambda \over \mu })^n}},\ee
where $\mu = 1-\lambda $.
We call this model {\it Generalized Totally Asymmetric Exclusion
Process}. In the limit $\lambda \rightarrow 0$, we have
$r_n=1, \ \ \forall n$, and in the limit $\lambda =1$, we have
 $r_0=1$ and $r_{n\ne 0}=0$. Note also that $r_{n+1}\leq r_n
,\ \ \forall n$. Therefore this process is perhaps more physical
than the two extreme cases studied in [1] and in section 2 of this paper.

In section 4, we use the coordinate Bethe ansatz and solve the master
equation of the process defined in section 3,
 and show that there is no bound state in the spectrum.

In section 5, we write the quantum Hamiltonian formalism for the
generalized process and, using a particle-hole exchange transformation,
show that this generalized process is equivalent
(i.e. in the same universality class ) to another process,
where particles hop only to the left. In this new process, if a
left neighbouring site is occupied,
 the move is rejected, but if a set of the left neighbouring sites are
empty, the particle hops with distance dependent rates to these sites:
$$ 0A \rightarrow A0 \hskip 2cm {\rm with} \ \ {\rm  rate } \ \ r_0 
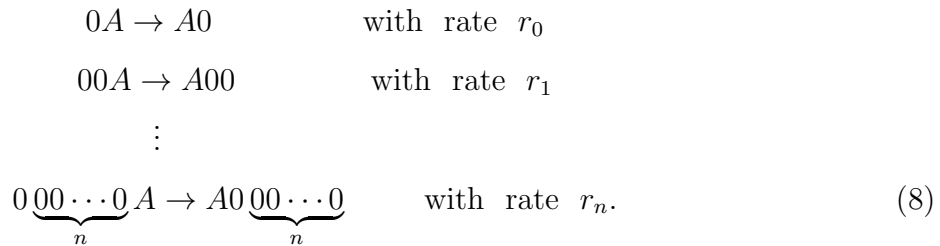
$$
$$ 00A \rightarrow A00 \hskip 1.8cm {\rm with} \ \ {\rm rate } \ \ r_1 $$
$$ \vdots \hskip 4.2cm $$
\be 0\underbrace {00\cdots 0}_nA \rightarrow A 0\underbrace {00\cdots 0}_n
\hskip 1cm {\rm with}\ \ {\rm rate } \ \ r_n. \ee
Therefore a transformation as simple as a particle-hole exchange, when
applied to our generalized process, has an interesting physical
consequence. Models with different values of $\la$,
 all allow exact solutions in the form of coordinate Bethe ansatz
 and their spectrum have only the continuous part, but only the limiting
cases of these models ($\lambda =0$ and $1$ ) allow a closed solution
in the form of a determinant.\\
We end up the paper with conclusion in section 5.

\section{ Generalized totally asymmetric exclusion process with $\la =0$ }

We augment the master equation (1) with the boundary condition (3).
Although we derive the rates
for arbitrary $\la $ in the next section by a general argument,
here we want to show that for $\la =0$ case, the master equation (1)
(together with the boundary condition (3))
describe an $n=1$ drop-push dynamics. For simplicity, consider the
two particle
sector of $ n=1$ drop push-dynamics. The master equations are
\be {\partial\over \partial t}  P( k_1, k_2,t) = P( k_1-1 , k_2,t )
+ P( k_1, k_2-1,t ) - 2 P( k_1, k_2,t ), \hskip 1cm k_2>k_1+1, \ee
\be {\partial\over \partial t}  P( k, k+1,t) = P( k-1 , k+1,t )
+ P( k-1, k,t ) - 2 P( k, k+1,t ) . \ee
Now, if we choose the boundary condition
$$ \hskip 4.3cm P(k,k,t) = P(k-1,k,t), \hskip 1cm \forall k,
\hskip 5.3cm  (3) $$
eq. (10) can be written as
\be {\partial\over \partial t}  P( k, k+1,t) = P( k-1 , k+1,t )
+ P( k, k,t ) - 2 P( k, k+1,t ) , \ee
which is of the same form as (9). \\
In the three particle sector, the extra equation which needs to be
taken into account is
\be {\partial\over \partial t}  P( k, k+1,k+2) = P( k-1 , k+1,k+2 )
+ P( k-1, k,k+2 )+P(k-1,k,k+1) - 3 P( k, k+1,k+2 ) . \ee
Using the boundary condition (3), the second and the third terms on
the r.h.s.
of (12) can be written as
\be P(k-1,k,k+2)=P(k,k,k+2), \ee
\be P(k-1,k,k+1)=P(k,k,k+1)=P(k,k+1,k+1), \ee
which means that eq. (12) is equivalent to the following standard form
\be {\partial\over \partial t}  P( k, k+1,k+2) = P( k-1 , k+1,k+2 )
+ P( k, k,k+2 )+P(k,k+1,k+1) - 3 P( k, k+1,k+2 ) . \ee
This procedure can be repeated for any sector. We will give a general
proof in the
next section.

To solve the master equation (1), and the boundary condition (3), for the
conditional probability
$P(x_1,x_2,\cdots ,x_N;t|y_1,y_2,\cdots ,y_N;0)$, we set, following
Sch\"utz [1],
\be P(x_1,x_2,\cdots ,x_N;t|y_1,y_2,\cdots ,y_N;0)=e^{-Nt}{\rm det }
G_N, \ee
where $G_N$ is an $N\times N$ matrix with entries $G_{ij}=g_{i-j}
(x_i-y_j,t)$.
The functions $g_p(x,t)$ are to be determined. Writing $G_N$ as
\be G_N= {\rm det}\left[ \begin{array}  {c} G_1(x_1,t) \\ \vdots \\
G_i(x_i,t) \\ \vdots
\\ G_N(x_N,t) \end{array} \right] ,\ee
where
\be G_i(x_i)=(g_{i-1}(x_i-y_1,t),g_{i-2}(x_i-y_2,t),\cdots ,g_{i-N}
(x_i-y_N,t)) , \ee
and inserting (16) in (1), we obtain
\be \sum_{i=1}^N  {\rm det}
 \left[ \begin{array}  {c} G_1(x_1,t) \\ \vdots \\
{\partial \over \partial t}G_i(x_i,t) \\ \vdots
\\ G_N(x_N,t) \end{array} \right] =\sum_{i=1}^N{\rm det}
 \left[ \begin{array}  {c} G_1(x_1,t) \\ \vdots \\ G_i(x_i-1,t) \\ \vdots
\\ G_N(x_N,t) \end{array} \right] ,\ee
the solution of which is
\be {\partial \over \partial t}G_i(x_i,t)=G_i(x_i-1,t). \ee
Inserting (16) in the boundary condition (3), we obtain
\be {\rm det}\left[ \begin{array}  {c} G_1(x_1,t) \\ \vdots \\
G_{k-1}(x,t) \\ G_k(x,t) \\
\vdots \\ G_N(x_N,t) \end{array} \right] ={\rm det}
 \left[ \begin{array}  {c} G_1(x_1,t) \\ \vdots \\ G_{k-1}(x-1,t) \\
G_k(x,t) \\
\vdots \\ G_N(x_N,t) \end{array} \right] ,\ee
the solution of which is
\be G_{k-1}(x,t)=G_{k-1}(x-1,t)+\beta G_k(x,t) ,\ee
where $\beta $ is an arbitrary parameter. The explicit form of the function
$g_p(x,t)$ can now be determined: these functions, as seen by
eqs. (20) and (22), should
satisfy the following relations
\be {\partial \over \partial t}g_p(n,t)=g_p(n-1,t), \ee
\be g_p(n,t)=g_p(n-1,t)+\beta g_{p+1}(n,t).\ee
Defining the generating functions (or $z$-transforms)
${\tilde g}_p(z,t):=\sum_{n=- \infty}^\infty z^ng_p(n,t)$,
eqs. (23) and (24) are converted to
\be {\partial \over \partial t}{\tilde g}_p(z,t)=z{\tilde g}_p(z,t), \ee
and
\be {\tilde g}_{p+1}(z,t)={1 \over \beta }(1-z){\tilde g}_p(z,t), \ee
the solution of which is simply obtained as
\be {\tilde g}_p(z,t)=e^{zt}{\tilde g}_p(z,0)=e^{zt}
({ {1-z} \over \beta })^p{\tilde g}_0(z,0).\ee
${\tilde g}_0(z,0)$ is nothing but the generating function for
$g_0(n,0)$, the one
particle sector probabilities at $t=0$. Since $P(x,0|y,0)=g_0(x-y,0)=
\delta_{x,y}$,
we have $g_0(n,0)=\delta_{n,0}$, and thus ${\tilde g}_0(z,0)=1$,
giving finally
\be {\tilde g}_p(z,t)=e^{zt}({{1-z}\over \beta })^p. \ee
The parameter $ \beta $, as long as it is nonzero, drops out of the
determinant
and so we can set it equal to unity. The functions $g_p(n,t)$ are
obtained by expanding the generating
functions. Note that the functions ${ \tilde g}_p(z,t)$ should be
expanded in terms of positive powers of $z$, if $p<0$. This is due to the
fact that, for $p<0$, as $n\rightarrow -\infty$, the function
$g_p(n,t)$ tend
to zero, since this limit is in the physical region. This expansion
yields, formally,
\be g_p(n,t)=\sum_{k=- \infty }^n {p\choose{n-k}}
 {(-1)^{n-k} \over k!}t^k .\ee
If $ p \geq 0$, $g_p(n,t)$ is converted to a finite sum
\be g_{p\geq 0}(n,t)=\sum_{k=0}^{{\rm min}(n,p)} {p\choose k}
 {(-1)^k \over (n-k)!}t^{n-k} .\ee
In particular,
\be g_0(n,t)={ t^n \over n! }. \ee
If $ p< 0$, $g_p(n,t)$ is converted to another finite sum
\be g_p(n,t)=\sum_{k=0}^n{{ |p|+k-1}\choose{\vert p\vert -1}}
 {t^{n-k} \over (n-k)!} .\ee
We have thus obtained an explicit relation for the conditional
probability.

\section{ Generalized totally asymmetric exclusion process with
arbitrary $\la$ }

We now consider the master equation (1) together with the boundary
condition
$$\hskip 3.5cm P(k,k,t) =\lambda P(k,k+1,t)+\mu P(k-1,k,t),
\hskip 1cm \forall k
.\hskip 2.8cm (5) $$
It can be easily shown that the conservation of probability demands
that $\mu =1-\la $.
In order to understand what type of process is described
by these equations, we first look at the two particle case.
Eqs. (1) and (5) yield
\begin{eqnarray} {\partial\over \partial t}  P( k, k+1)&=& P( k-1 , k+1 )
+ P( k, k ) - 2 P( k, k+1 )\cr
&=&P(k-1,k+1)+\mu P(k-1,k)-(1+ \mu )P(k,k+1) ,\end{eqnarray}
which means the following rates.
$$ A0 \rightarrow 0A \hskip 2cm {\rm with} \ \ {\rm  rate } \ \ r_0=1, $$
$$ AA0 \rightarrow 0AA \hskip 1.8cm {\rm with} \ \ {\rm rate } \ \
r_1=\mu . $$
To find the rates in the general case, we first prove a lemma.\\
{\bf Lemma: } Equation (5) implies, for arbitrary $n$, the following
$$ P(k,k+1,k+2,\cdots ,k+n-1,k+n,k+n)=(1-r_{n+1})P(k,k+1,k+2,\cdots
,k+n-1,k+n,k+n+1) $$
\be +r_{n+1}P(k-1,k,k+1,\cdots ,k+n-2,k+n-1,k+n), \ee
where
\be r_n=(1+{\la \over \mu}+({\la \over \mu})^2+\cdots +
({\la \over \mu})^n)^{-1} .\ee
{\bf Proof: } We proceed by induction. For $n=0$, eqs. (34) and (35)
are the
same as (5), as $r_1=\mu$. Assuming now, that (34) and (35) are
correct for
$n-1$, and using eq. (5), we have
$$ P(k,k+1,\cdots ,k+n-1,k+n,k+n)$$
$$=\la P(k,k+1,\cdots ,k+n-1,k+n,k+n+1)
+\mu P(k,k+1,\cdots ,k+n-1,k+n-1,k+n) $$
$$=\la P(k,k+1,\cdots ,k+n-1,k+n,k+n+1)$$
\be +\mu \{(1-r_n) P(k,k+1,\cdots ,k+n-1,k+n,k+n)
 +r_nP(k-1,k,\cdots ,k+n-2,k+n-1,k+n) \} , \ee
or
$$ P(k,k+1,\cdots ,k+n-1,k+n,k+n)=s_{n+1}P(k,k+1,\cdots ,k+n-1,k+n
,k+n+1) $$
\be +r_{n+1}P(k-1,k,\cdots ,k+n-2,k+n-1,k+n), \ee
where
\be {\la \over 1-\mu (1-r_n)}=s_{n+1} \hskip 0.5cm , \hskip 0.5cm
{\mu r_n \over 1-\mu (1-r_n)}=r_{n+1}. \ee
From eq.(38), it is seen that $s_{n+1}+r_{n+1}=1$. One
can now solve the second equation of (38) for $r_{n+1}$ to obtain
$$ {\mu r_n \over \la+\mu r_n}=r_{n+1} \hskip 0.5cm  {\rm or}
\hskip 0.5cm r_{n+1}^{-1}={\la \over \mu} r_n^{-1}+1, $$
which gives
$$r_{n+1}^{-1}=1+{\la \over \mu}\{1+{\la \over \mu}+({\la \over \mu})^2
+\cdots +({\la \over \mu})^n \}=
1+{\la \over \mu}+({\la \over \mu})^2+\cdots +({\la \over \mu})^{n+1}.$$
This proves the lemma.

We now consider a collection of $n$ adjacent particles and write the
master equation
for this configuration by eq. (1):
$$ {\partial  P\over \partial t} (k,k+1,k+2,\cdots ,k+n-1)= $$
\be \sum_{i=0}^{n-1}P(k,k+1,\cdots ,k+i-2,k+i-1,k+i-1,k+i+1,\cdots
,k+n-1) -nP(k,k+1,k+2,\cdots ,k+n-1). \ee
Using (34), we find
$$ {\partial  P\over \partial t} (k,k+1,k+2,\cdots ,k+n-1)=
\sum_{i=0}^{n-1}r_iP(k-1,k,\cdots ,k+i-2,k+i-1,k+i+1,\cdots ,k+n-1)$$
\be - \left( \sum_{i=0}^{n-1}r_i \right) P(k,k+1,k+2,\cdots ,k+n-1).\ee
It is now obvious that the above equation describes a process in which a
collection of $i+1$ adjacent particles hop to the right with rate $r_i$,
as claimed in the introdution.

\section{ The Bethe ansatz solution for arbitary $\la$ }

In this section we denote the position of the particles by
$x_i \in {\bf Z} $ rather than $k_i$,
and apply the Bethe ansatz for the solution of the master equation (1)
and the boundary condition (5). Writing $P_N(x_1,\cdots ,x_N,t)=e^{-\ep_Nt}
\Psi_N(x_1,\cdots ,x_N)$, will turn (1) into an eigenvalue equation for
$ \Psi_N(x_1,\cdots ,x_N)$:
\be \sum_{i=1}^N\Psi_N(x_1,\cdots ,x_i-1,\cdots ,x_N)=(N-\ep_N)
\Psi_N(x_1,\cdots ,x_i,\cdots ,x_N). \ee
We write the coordinate Bethe ansatz for $\Psi $ in the form:
\be \Psi_N(x_1,\cdots ,x_N)=\sum_\sigma A_\sigma e^{i \sigma ({\bf p})
.{\bf x}}, \ee
where ${\bf x}$ and ${\bf p} $ stand for the $n$-tuple coordinates and
momenta
and $\s ({\bf p}) $ is a permutation of momenta. The sum is over all
permutations.
Inserting (42) into (41) yields
\be \sum_\sigma A_\sigma e^{i \sigma ({\bf p}).{\bf x}}\left( e^{-i\s
(p_1)}
+e^{-i\s (p_2)}+ \cdots + e^{-i\s (p_N)} \right) =(N-\ep_N )\Psi_N(x_1
,\cdots
,x_N). \ee
The sum in the paranthesis can be written as $\sum_{k=1}^Ne^{-ip_k}$ and
taken outside $\sum_\s $, yielding
\be \ep_N:=\sum_{k=1}^N\ep (p_k)=\sum_{k=1}^N(1-e^{-ip_k}).\ee
Note that due to translational invariance, $\Psi_N$ is also an eigenvector
of total momentum $P$, which in the lattice is defined as the logarithm
of the shift operator $U=e^{-iP }$:
\be (U\Psi_N)(x_1,\cdots ,x_N):=\Psi_N(x_1-1,x_2-1,\cdots ,x_N-1).\ee
Acting by $U$ on (42), we obtain
\be (P \Psi_N)(x_1,\cdots ,x_N)=(p_1+\cdots +p_N)\Psi_N(x_1,\cdots ,
x_N). \ee
Therefore the eigenvectors $\Psi_N$ have additive total {\it energies}
 and momenta.
Inserting (42) in the boundary condition (5), rewritten in an
unabbreviated form:
$$ \Psi (x_1,\cdots ,x_i=\xi ,x_{i+1}=\xi ,\cdots , x_N)=\la
\Psi (x_1,\cdots ,x_i=\xi ,x_{i+1}=\xi +1 ,\cdots , x_N) $$
$$ + \mu\Psi (x_1,\cdots ,x_i=\xi -1 ,x_{i+1}=\xi ,\cdots , x_N), $$
we obtain
\be \sum_\s e^{i\sum_{k \not = i,i+1}\s (p_k )x_k+i(\s (p_i)+\s
(p_{i+1}))\xi }
\left[ A_\s \left( 1-\la e^{i\s (p_{i+1})}-\mu e^{-i\s (p_i)} \right)
 \right]=0. \ee
We denote the expression in the bracket by $B_\s $. Noting that the
prefactor is
unaffected by an interchange of $p_i$ and $p_{i+1}$, it follows that
 the proper
coefficient of each prefactor, which should vanish, is
$B_\s +B_{\s \s_i} $, where $\s_i$ is the generator of $S_N$ (the
permutation group of
$N$ object ) which only interchanges $ p_i$
and $ p_{i+1} $
\be \s_i(p_1,\cdots ,p_i, p_{i+1},\cdots  ,p_N)=(p_1,\cdots
,p_{i+1}, p_i,\cdots
,p_N), \ee
and $\s \s_i $ stands for the product of two group elements, $ \s $
acting after
$\s_i$. Therefore we find:
$$A_{\s}(1-\lambda e^{i\s(p_{i+1})} - \mu e^{-i\s(p_{i})})
+ A_{\s \s_i}(1-\lambda e^{i\s(p_{i})} - \mu e^{-i\s(p_{i+1})})= 0 ,$$
or
\be {A_{\s \s_i}\over A_{\s} } = { \lambda e^{i\s(p_{i+1})} + \mu
e^{-i\s(p_{i})}-1\over { 1-\lambda e^{i\s(p_{i})} - \mu
e^{-i\s(p_{i+1})}}}
= S(\s(p_i), \s(p_{i+1})).  \ee
This relation, in effect, allows one to find all the $ A_{\s} $'s in
terms of
$ A_1$ (which is set to unity). The first few coefficients, corresponding
to the elements $ 1, \s_1, \s_2, \s_1 \s_2, \s_2\s_1, \s_1\s_2\s_1 $ are :
$$ A_1 = 1 \hskip 1cm ,  \hskip 1cm A_{\s_1} = S_{12} \hskip 1cm ,
\hskip 1cm
A_{\s_2} = S_{23}, $$
\be A_{\s_1\s_2} = S_{12}S_{13} \hskip 1cm ,  \hskip 1cm
 A_{\s_2\s_1} = S_{13}S_{23} \hskip 1cm ,  \hskip 1cm A_{\s_1\s_2\s_1} =
S_{12}S_{13}S_{23}, \ee
where $S_{ij} = S(p_i,p_j) $.
The form of the scattering matrix $S_{ij} $ could also be found from the
two particle sector alone. The above analysis shows in fact the
factorizibility
of the $S$ matrix in the general case, a sign of the integrabilty of
the problem.

To find the range of $ p_i $'s, we analyze the $S$ matrix
\be S_{12} = { \la e^{ip_2} + \mu e^{-ip_1}-1 \over  1-\la e^{ip_1} - \mu
e^{-ip_2}} = {c_{21}\over c_{12}}, \ee
and the two particle wave function
$$\Psi_2(x_1, x_2) = c_{12} e^{i(p_1x_1+p_2x_2)} + c_{21}
e^{i(p_2x_1+ p_1x_2)}, $$
or
\be \Psi(X,x) = e^{iPX}\left( c_{12} e^{ipx} + c_{21} e^{-ipx} \right), \ee
where $X:={1\over 2}( x_1 + x_2) $, $x:= x_1 - x_2 $ , $P:= p_1 + p_2 $ and
$p:= {1\over 2} ( p_1 - p_2 ) $ with clear physical meanings.
Since $ x $ is negative ( $ x_1 < x_2 $ ), to have a bound state one
of the following set of condintions should be satisfied simultaneously.
Either
$$ c_{12} = 0 \hskip 1cm ,  \hskip 1cm  {\rm Im}p> 0 \hskip 1cm ,
\hskip 1cm {\rm Im}P = 0 ,$$
or
$$ c_{21} = 0 \hskip 1cm ,  \hskip 1cm  {\rm Im}p< 0 \hskip 1cm ,
\hskip 1cm {\rm Im}P = 0 .$$
Rewriting $ c_{12} $ in terms of the new momenta, we find
\be c_{12} = e^{ip} ( e^{-ip} - \la e^{i{P\over 2}} - \mu
e^{-i{P\over 2}}). \ee
Since Im$P=0 $  we have
\be \vert \la e^{i{P\over 2}} +\mu e^{-i{P\over 2}}\vert \leq \la
+ \mu =1. \ee
Noting that $\vert e^{-ip} \vert >1 $, it is seen that $ c_{12} $
can not vanish.
A similar analysis applies for the second set of conditions.
Therefore no bound
state exists in the spectrum, and the range of all momentum
variables is $[0, 2\pi)$.

To find the conditional probability
$P_N( x_1, x_2,..., x_N ;t \vert y_1, y_2, ... y_N;0) $, one
should take a
linear combination of the eigenfunctions $\Psi_N$, with suitable
coefficients.
Consider the two particle sector. We  have
\be P_2( x_1, x_2;t \vert y_1, y_2; 0) = \int_{0}^{2\pi} \int_{0}^{2\pi}
{dp_1\over 2\pi}{dp_2\over 2\pi}
e^{-[\epsilon (p_1) + \epsilon (p_2)]t - ip_1y_1-ip_2y_2}
\Psi_2(x_1,x_2). \ee
This is just a linear combination of the eigenfunctions,
satisfying the initial condition\\
$$P_2( x_1, x_2;0 \vert y_1, y_2; 0)=
\delta_{x_1,y_1}\delta_{x_2,y_2},$$
in the physical region ($x_2>x_1,\;\;
y_2>y_1$). The eigenfunction $\Psi_2(x_1, x_2)$ in (55) is normalized
according to
$$\Psi_2(x_1, x_2) = e^{i(p_1x_1+p_2x_2)} + S_{12} e^{i(p_2x_1+
p_1x_2)}. $$
To avoid the singularity in $S_{12}$, we set $p_1\to p_1+i\epsilon$.
With this
prescription, the contribution of the second term in $\Psi_2$ to
$P_2(x_1, x_2, 0\vert y_1, y_2, 0)$ identically vanishes in the physical
region. Using the variables $\xi :=e^{ip_1}$ and $\eta :=e^{-ip_2}$,
a simple
contour integration yields
\begin{eqnarray} P_2 ( x_1, x_2 ; t \vert y_1, y_2; 0 )&=&e^{-2t}\left\{
{{t^{x_1-y_1}}\over{(x_1-y_1)!}}{{t^{x_2-y_2}}\over{(x_2-y_2)!}}\right.\cr
&& -\sum_{k=0}^{\infty}\sum_{m=0}^k{k\choose m}\la^m\mu^{k-m}
{{t^{x_2-y_1+m}}\over{(x_2-y_1+m)!}}{{t^{x_1-y_2-k+m}}
\over{(x_1-y_2-k+m)!}}
\cr &&\times\left.\left[ 1-{{\la t}\over{x_1-y_2-k+m+1}}-
{{\mu (x_2-y_1+m)}\over t} \right]\right\} .
\end{eqnarray}
It is easy to see, explicitly, that this solution satisfies the initial
condition in the physical region. Also, in the limiting cases $\la =1$ and
$\la =0$, it reduces, respectively, to
\begin{eqnarray} P_2 ( x_1, x_2 ; t \vert y_1, y_2; 0 )&=&e^{-2t}\left\{
{{t^{x_1-y_1}}\over{(x_1-y_1)!}}{{t^{x_2-y_2}}\over{(x_2-y_2)!}}\right.\cr
&&\left. -\left[
{{t^{x_1-y_2}}\over{(x_1-y_2)!}}-{{t^{x_1-y_2+1}}\over{(x_1-y_2+1)!}}
\right] \sum_{k=0}^{\infty}{t^{x_2-y_1+k}\over{(x_2-y_1+k)!}}
\right\} ,
\end{eqnarray}
obtained in [1], and
\begin{eqnarray} P_2 ( x_1, x_2 ; t \vert y_1; y_2, 0 )&=&e^{-2t}\left\{
{{t^{x_1-y_1}}\over{(x_1-y_1)!}}{{t^{x_2-y_2}}\over{(x_2-y_2)!}}\right.\cr
&&\left. -\left[
{{t^{x_2-y_1}}\over{(x_2-y_1)!}}-{{t^{x_2-y_1-1}}\over{(x_2-y_1-1)!}}
\right] \sum_{k=0}^{\infty}{t^{x_1-y_2-k}\over{(x_1-y_2-k)!}}
\right\} ,
\end{eqnarray}
obtained in the present paper.

The treatment of the $N$ particle case is similar. We have
$$ P_N( x_1, \cdots , x_N;t \vert y_1, \cdots , y_N; 0) =
\int_{0}^{2\pi}{dp_1\over 2\pi}\cdots \int_{0}^{2\pi}{dp_N\over 2\pi}
e^{-[\sum\epsilon (p_i)]t - i\sum p_iy_i}\Psi_N(x_1, \cdots , x_N). $$
The integration is defined with the following $\epsilon$-prescription: in
$S_{ij}$ ($i<j$), $p_i$ is replaced by $p_i+i\epsilon$.

\section{ Hamiltonian approach}

The Hilbert space of generalized totally asymmetric exclusion process is
$ {\cal H} = \otimes C_2$
, the tensor product of all the local Hilbert spaces of the lattice sites.
 $ C_2 $ is the two dimensional vector space with basis states
$ |0>= \left( \begin{array}  {c} 1 \\ 0 \end{array} \right)$ and
$ |1>= \left( \begin{array}  {c} 0 \\ 1 \end{array} \right)$.
The states $|0>$ and $ |1>$  represent vaccant and occupied
sites, respectively. The local operators $ n_i= \left( \begin{array}
 {cc} 0&0 \\ 0&1
\end{array} \right)$, $ \s_i^+= \left( \begin{array}  {cc}
0&1 \\ 0&0 \end{array} \right) $, and $ \s_i^-= \left( \begin{array}
 {cc} 0&0 \\ 1&0 \end{array} \right) $ are  the number
, annihilation, and creation operators, respectively. Their action on
a bra state
$ < \alpha \vert$, ($\alpha = 0, 1 )$, can be conveniently represented as
$  < \alpha \vert n = \alpha < \alpha \vert$, $< \alpha \vert \s^+
= ( 1-\alpha)
< 1-\alpha \vert $ and $< \alpha \vert \s^-= \alpha <1-\alpha \vert $ .

The state of the system $ \vert \Psi (t)> $ evolves according to the
Schr\"odinger type
equation $ -{ \partial \over \partial t} \vert \Psi(t)> = H \vert
\Psi (t) > $.
The connection between the two representations is given by the relation
\be P(k_1, k_2,\cdots ,k_N ,t) = < k_1, k_2,\cdots ,k_N \vert \Psi_N(t)
>=<0\vert \s^+_{k_1} \s^+_{k_2}\cdots ,\s^+_{k_N} \vert \Psi_N (t) > .\ee
The Hamiltonian of the process can now be
written as
\be H = -\sum_{k\in L } \sum_{l\geq 1} r_{l-1} ( v_k(l) - w_k(l) ), \ee
where $L$ represents the sites of the lattice and
\be  v_k(l) = \s^+_kn_{k+1} n_{k+2}\cdots n_{k+l-1}\s^-_{k+l},\ee
\be  w_k(l) = n_kn_{k+1}n_{k+2}\cdots n_{k+l-1}(1-n_{k+l}).\ee
Consider a bra state containing $l$ particles on adjacent sites:
$ < k+1, k+2,\cdots ,k+l \vert $.  The only terms in $ H $ with
nonvanishing action on this state are
\be < k+1, k+2,\cdots ,k+l \vert v_k(i) = < k, k+1,\cdots
,k+i-1,k+i+1,\cdots ,k+l \vert ,  \;\; 1\leq i \leq l, \ee
\be < k+1, k+2,\cdots ,k+l \vert w_{k+1+l-i}(i) = < k+1,
k+2,\cdots ,k+l \vert ,  \;\; 1 \leq i \leq l .\ee
Note that the action of the above operators on every other
state which contains, beside the above particles, other collection of
particles disconnected from the above one is the same. Using
eqs. (59) to (64),
one arrives at eq. (40) for the evolution of the probability.
Note that the quantum Hamiltonian (60) is a stochastic
operator, meaning that all of it's off-diagonal matrix elements
are non-positive
with the sum of entries in each column being equal to zero. This
last property
is expressed by saying that $ < S \vert H = 0 $ where
$ < S \vert $ is the sum
of all basis states of $ {\cal H } $. Equivalent models may be obtained
by constructing operators
$\Omega : {\cal H }\longrightarrow {\cal H } $ and Hamitonians
$ H'= \Omega H \Omega^{-1} $, which preserve the above properties.
An obvious example is the particle-hole exchange operator
$ \Omega = \prod_i {\bf \s}^x_i$. It
clearly has the property that $ < S \vert \Omega = < S \vert $, so that
for $ H ' $ we also have $ < S \vert H' = 0  $. It is easy to see that the
this transformation induces the changes $n\leftrightarrow 1-n$ and
$\sigma^+\leftrightarrow \sigma^-$. So the
master equation obtained from $H'$ describes the process (8).

\section{ Discussion and Outlook}

We have defined a generalized exclusion process, parameterized by a
real parameter
$ \lambda \in [0,1] $, and have shown that the master equation of
this model admits
for every $ \lambda $ an exact solution via the coordinate Bethe ansatz.
 We have also shown that this
model interpolates continously between two very different models:
the totally asymmetric exclusion model (for $ \lambda =1 $), which we may
consider as the weakcoupling limit and the drop-push model (for
$ \lambda = 0 $),
which may be considered as the strong coupling limit of the model.
In these two limits, the solution acquires a simple determinant form.

 Our work can be further investigated in one definite way. It may be
that the point $ \lambda = {1\over 2} $ is a point of phase transition
and the
study of the equilibrium properties of the model on a periodic lattice
may reaveal this transition. There are already two pieces of evidence for
the validity of this conjecture. First, there is some sort of duality
between
two models two models with parameters symmetric with respect to
$\la =1/2$. To be more specific, we have
\be S(p_1, p_2;\la ) =   S(-p_2, -p_1;1- \la ). \ee
Second, the large $l$ behaviour of the transition rates is
$$  r_l\sim
\left\{ \begin{array}{ll} 1-{\la \over \mu}, &  \mbox{$ \la
< {1\over 2}$} \\  {1\over l}, & \mbox{$\la ={1\over 2}$}\\
\left({\la\over\mu}\right)^{-l},  & \mbox{$ \la > {1\over 2}$}
\end{array}  \right. . $$
It will be interesting to study the stationary behaviour of this
system along the lines
which have been followed in [11-14], to see what kind of phases develop
in the system by varying $\la $.

\vskip 1cm

\noindent{\bf Acknowledgement}

V. K. would like to thank M. E. Fouladvand for useful
discussions.
M. Alimohammadi would like to thank the research council of the Tehran
university, for partial financial support.

\end{document}